\documentclass[a4paper,12pt]{article}
\usepackage{epsfig,amsmath,amsfonts}
\usepackage[T2A]{fontenc}            % ?????????? ?????????  TeX
\usepackage[utf8]{inputenc}
\newcommand{\eqsection}{\makeatletter
   \@addtoreset{equation}{section}
   \renewcommand{\theequation}{\arabic{section}.\arabic{equation}}
   \makeatother}
\def\ear{\end{eqnarray}}
\def\beq{\begin{equation}}             \def\earn{\nonumber \end{eqnarray}}
\def\eeq{\end{equation}}               
\def\bear{\begin{eqnarray}}            
\def\Half{{\frac{1}{2}}}

\def\agb{\bar{\alpha}}

\begin{document}

\begin{center}
\LARGE{\bf {New Exact Solutions for a Chiral Cosmological Model in 5D EGB Gravity}} %%
\end{center}

\begin{center}

Maharaj S.D., \\%%

\sl{\it {Astrophysics and Cosmology Research Unit \\
School of Mathematics, Statistics and Computer Science,\\
University of KwaZulu-Natal, Private Bag X54 001 \\
Durban 4000, South Africa}\\
Email: maharaj@ukzn.ac.za }
\bigskip

Beesham A., \\%%

{\it {Department of Mathematical Sciences,  University of Zululand \\
Private Bag X1001, Kwa-Dlangezwa 3886, South Africa }\\
Email: beeshama@unizulu.ac.za }

\bigskip
Chervon S.V.,\\%

\sl{\it {Astrophysics and Cosmology Research Unit \\
School of Mathematics, Statistics and Computer Science, \\
University of KwaZulu-Natal, Private Bag X54 001 \\
Durban 4000, South Africa}} and

\sl{\it {Laboratory of Gravitation, Cosmology, Astrophysics,\\
Ulyanovsk State Pedagogical University, \\100-years V.I. Lenin's Birthday square, Ulyanovsk 432071, Russia}
Email: chervon.sergey@gmail.com }

\bigskip
Kubasov A.S. \\
\sl{\it {Laboratory of Gravitation, Cosmology, Astrophysics,\\
Ulyanovsk State Pedagogical University, \\100-years V.I. Lenin's Birthday square, Ulyanovsk 432071, Russia}
Email: as-kubasov@rambler.ru }

\end{center}

\bigskip

\small{We consider a chiral cosmological model in the framework of Einstein-Gauss-Bonnet cosmology. Using a decomposition of the latter equations in such a way that the first chiral field is responsible for the Einstein part of the model, while the second field together with the kinetic interaction is connected with the Gauss--Bonnet part of the theory, we find new exact solutions for the 2-component chiral cosmological model with and without the kinetic interaction between fields.
}

\bigskip

\small{Мы рассматриваем Киральную Космологическую Модель (ККМ) в рамках космологии ЭГБ. Используя разбиение уравнений ЭГБ таким образом, что первое киральное поле ответственно за эйнштейновскую часть модели, в то время как второе поле, вместе с кинетическим взаимодействием, связано с вкладом теории Гаусса-Бонне. Мы нашли новые точные решения для 2-х компонентной ККМ с кинетическим взаимодействием между полями и без него.
}

\newpage

\section{Introduction}

The form of the potential of self-interaction for scalar field theory in cosmology is an interesting topic.
The difference between scalar potentials in particle physics and those in cosmology has been stressed in the work \cite{Halliwell:1986ja}. Halliwell wrote "... we do not really know which theory of particle physics best describes the very early universe. One should therefore keep an open mind as to the form of $V(\phi)$."

To find exact solutions, usually one suggests that we know from high energy physics the scalar potential in the very early universe, and our task is to find the scale factor and the scalar field as  functions of time. The work by Ellis and Madsen \cite{Ellis:1990wsa} was the first one to consider "the inverse problem" within the framework of cosmology.  These authors \cite{Ellis:1990wsa} suggested to start from a given scale factor instead! Indeed, it is clear that the scale factor may be found from observational data. Then we may take into account this fact to find the potential and scalar field from the cosmological equations. This work was done and examples of exact solutions have been presented for pure scalar fields (without taking into account radiation which was also considered there). Further, this approach was developed in the works \cite{Chervon:1996ju, Chervon:1997yz}. In our study, we will use such an approach, the so-called "fine tuning of the potential method", to find new solutions in cosmology based on Einstein-Gauss-Bonnet (EGB) gravity in five-dimensional (5D) spacetime.

%%%%%%%%
The feature of our approach is in the fact that we use the chiral cosmological model (CCM) \cite{Chervon:2014dya} reduced to a two dimensional target space. That is, for our consideration in the present contribution, we consider two scalar fields with or without the kinetic interaction term. Such an  approach has been considered, for example, with the aim of exact solutions construction in \cite{Vernov:2006tmf,Paliathanasis2014prd} and for studying  inflation in f(R) gravity in \cite{Bamba:2015epj}. The set of exact solutions in the 2-component CCM for 5D EGB gravity  for the Emergent Universe was found in the work \cite{Chervon:2014tra}. The evolution of the scale factor was taken in the form $a(t)=A\left(\beta+e^{\alpha t}\right)^m$. Also in \cite{Chervon:2014tra}, it was mentioned about the possibility to extend the method to exponential, power-law and other kinds of  solutions.

In the present investigation, we use once again the method of the decomposition of the field-gravity equations and  the definition  of special ansatzes. The method was described for GR in {\cite{Beesham:2014tja} and for the case of EGB gravity in \cite{Chervon:2014tra}. We study new possibilities for exact solution construction for various types of  evolutions, including power-law, power-law-exponential and hyperbolic scale factors. We stress that the obtained solutions are 5D ones, and that there is no way to compare exact solutions in 5D EGB with those in Frisdmann cosmology based on Einstein's theory. It is not possible to define cosmological parameters from 5D EGB cosmology without compactification of the fifth dimension. However, this is not the main aim of this paper, and will be a subject for future investigation.

\section{Basic equations of the model}

Chiral cosmological fields are considered as the source of gravitation in the EGB model. The action of the model is:
\begin{equation}\label{act-gen}
S=\int d^5x\sqrt{-g}\left(\frac{1}{2}R+\frac{1}{2}\agb R_{GB}+\frac{1}{2}h_{AB}(\varphi)\varphi^A_{,a}\varphi^B_{,b}g^{ab}-V(\varphi) \right),
\end{equation}
where $R_{GB}$ is the Lovelock tensor defined as $R_{GB}=R^2-4R_{ab}R^{ab}+R_{abcd}R^{abcd}$ and $\agb $ is the Gauss-Bonnet parameter. Other notations correspond to those in \cite{Beesham:2014tja}.

We will study cosmology in 5D space-time with the Friedmann-Robertson-Walker (FRW)-type metric
\begin{equation}\label{frw-5d}
dS^2=-dt^2+a(t)^2\left(\frac{dr^2}{1-\epsilon r^2}+r^2(d\theta^2+\sin^2\theta(d\varphi^2+\sin^2\varphi d\chi^2))\right),
\end{equation}
and we will take a CCM with the 2-component diagonal metric of a target-space
\begin{equation}\label{mts-dig}
ds_{ts}=h_{11}d\phi^2+h_{22}(\phi,\psi)d\psi^2,
\end{equation}
as the source of gravity.

The system of fields and Einstein's equations may be displayed in the following form \cite{Chervon:2014tra}:
\begin{equation}\label{fried-1}
H^2+\frac{\epsilon}{a^2}+\agb\left(H^2+\frac{\epsilon}{a^2}\right)^2=\frac{1}{6}\left(\frac{1}{2}\dot\phi^2+\frac{1}{2}h_{22}(\phi,\psi)\dot\psi^2+V(\phi,\psi)\right),
\end{equation}
\begin{equation}\label{fried-2}
\left[1+2\agb\left(H^2+\frac{\epsilon}{a^2}\right)\right]\left(\dot H-\frac{\epsilon}{a^2}\right)=-\frac{1}{3}\left(h_{11}\dot\phi^2+h_{22}(\phi,\psi)\dot\psi^2\right),
\end{equation}
\begin{equation}\label{phi-dyn}
h_{11}\ddot\phi+4Hh_{11}\dot\phi-\frac{1}{2}\frac{\partial h_{22}}{\partial\phi}\dot\psi^2+\frac{\partial V}{\partial\phi}=0,\, h_{11}=constant,
\end{equation}
\begin{equation}\label{psi-dyn}
h_{22}(\phi,\psi)\ddot\psi+\dot h_{22}(\phi,\psi)\dot\psi+4Hh_{22}(\phi,\psi)\dot\psi-\frac{1}{2}\frac{\partial h_{22}}{\partial\psi}\dot\psi^2+\frac{\partial V}{\partial\psi}=0.
\end{equation}

Our main attention will be to find exact solutions and we consider the case of the FRW spatially-flat universe ($\epsilon=0$). The kinetic energy and potential can be easily obtained from the basic equations  (\ref{fried-1})-(\ref{psi-dyn}) and read:
\begin{equation}\label{K-on-t}
-\frac{2}{3}K(t)=\left[1+2\agb H^2\right]\dot H=-\frac{1}{3}\left(h_{11}\dot\phi^2+h_{22}(\phi,\psi)\dot\psi^2\right),
\end{equation}
\begin{equation}\label{V-on-t}
\frac{V}{6}=H^2+\frac{1}{4}\dot H+\agb H^2\left(H^2+\frac{1}{2}\dot H\right).
\end{equation}
We may solve the system (\ref{fried-1})--(\ref{psi-dyn}) using the decomposition suggested in \cite{Chervon:2014tra}
\begin{equation}\label{antz-1}
h_{11}\dot\phi^2=-3\dot{H},\,~h_{22}(\phi,\psi)\dot\psi^2=-6\agb H^2
\dot H,\,~V(\phi,\psi)=V_1(\phi)+e^{f(\phi)}V_2(\psi),
\end{equation}
where
\begin{equation}\label{anz-V-H}
V_1(\phi(t))=6H^2+\frac{3}{2}\dot
H,\,~e^{f(\phi(t))}V_2(\psi(t))=6\agb H^2\left(H^2+\frac{1}{2}\dot
H\right).
\end{equation}

Here $f(\phi)$ is a function on time, defined from compatibility later on, and  we set $h_{11}$  equal to unity: $h_{11}=1$.
The relation between the metric coefficient $h_{22}$ and $f(\phi)$ is dependent on the law of evolution of the Hubble parameter and is determined from the equation:
\begin{equation}\label{f-h22}
\frac{\partial\ln(h_{22})}{\partial t}=-\frac{2H^2+\dot H}{\dot
H}\frac{\partial f}{\partial t}.
\end{equation}
For further investigation we will use the equation of state in the following form
\beq\label{EoS}
\omega=-1-\frac{\dot{H}}{2H^2}\left(\frac{1+2\agb H^2}{1+\agb H^2}\right).
\eeq

Thus we have one free parameter, namely, a kinetic interaction $h_{22}$ as a function of the fields. It is worthwhile to mention that it is difficult to obtain the kinetic interaction term from observations \cite{Abbyazov:2012pt}, but  numerically  the reconstruction of the target space metric component  $h_{22}$ and the potential of interaction $V(\phi,\psi)$ was carried out in the work \cite{Abbyazov:2013mpla}.

\section{Exact solutions without kinetic interaction between fields}

In the work \cite{Chervon:2014tra} it was proved that, using the decomposition (\ref{antz-1})-(\ref{anz-V-H}), exact solutions for the 2-component CCM for the emergent universe can be obtained only for chiral fields without kinetic interaction between them. In our present consideration, we will show that both cases are possible: without kinetic interaction and with it.
In the next section, we will present the solutions with kinetic interaction between the chiral fields. Here we will start with the case when the kinetic interaction between fields is absent.
For this purpose we set $h_{11}=h_{22}=1$. After that simplification we can represent the "ansatz solution" as:
\beq\label{phi-e1}
\dot{\phi}^2=-3\dot{H},
\eeq
\beq\label{psi-e1}
\dot{\psi}^2=-6\agb H^2\dot{H},
\eeq
\beq\label{V1-phi-e1}
V_1(\phi)=6H^2+\frac{3}{2}\dot{H},
\eeq
\beq\label{V2-psi-e1}
V_2(\psi)=6\agb H^2\left(H^2+\frac{1}{2}\dot{H}\right).
\eeq
Substitution of equations (\ref{phi-e1})-(\ref{V2-psi-e1}) into the field equations (\ref{phi-dyn})-(\ref{psi-dyn}) gives us identities.
Note that for $h_{22}=1$, we obtain $f(\phi)=const$, and the term $e^{f(\phi)}$ may be included in  $V_2(\psi)$.

Thus to construct exact solutions, we may use the evolutionary law of the universe. That is, for a given scale factor $a(t)$ (or Hubble parameter $H(t)$), we can obtain the potential and the kinetic energy as  functions of time. Chiral fields may be defined from equations (\ref{phi-e1})-(\ref{psi-e1}) in quadratures,  or in elementary functions depending on the given scale factor.

\subsection{Power-law evolution}

Power-law evolution is very important for a variety of reasons, e.g.,  such exact solutions  can solve the horizon, flatness and perturbation-spectrum problems in FRW cosmology.
Thus we choose the scale factor as
\beq\label{a-powlaw}
a(t)=At^m,~~A=const.,~~A>0, ~~m>1.
\eeq
Under these conditions above we have $H_{pl}=m/t$ and the acceleration of the Universe is always positive : $\frac{\ddot{a}}{a}=H^2+\dot{H}=\frac{m(m-1)}{t^2}>0.$

The solution for $\phi$ from (\ref{phi-e1}) is
\beq\label{phi-sol-1}
\phi-\phi_*=\pm \sqrt{3m}\ln t .
\eeq
Hereinafter the letters with a subscript-star mean constants of integration.
The second field $\psi$ can be defined by integrating (\ref{psi-e1}) and it gives

\beq\label{psi-sol-1}
\psi-\psi_*=\mp \frac{\sqrt{6\agb m^3}}{t}.
\eeq

Using the obtained solutions for the chiral fields $\phi$ and $\psi$, we can reconstruct the potentials $V_1$ and $V_2$ from their dependence on time $t$. Finally we obtain
\beq\label{V1-sol-1}
V_1(\phi)=\frac{3m}{2}(4m-1)\exp \left(\mp \frac{2(\phi-\phi_*)}{\sqrt{3m}}\right),
\eeq
\beq\label{V2-sol-1}
V_2(\psi)=\frac{2m-1}{12\agb m^3}\left(\psi-\psi_*\right)^4.
\eeq
The equation of state (\ref{EoS}) for the solution (\ref{phi-sol-1})-(\ref{V2-sol-1}) is
\beq\label{EoS-1}
\omega = -1+m^{-1}\left(1+\frac{\agb m^2}{t^2 +\agb m^2}\right).
\eeq

\subsection{Power-law-exponential evolution}

The scale factor for  power-law-exponential evolution of the universe is
\begin{equation}\label{plexp-at}
a(t)=At^m e^{\lambda t},~~\lambda=const.,~~\lambda>0,~~m>1.
\end{equation}
We come across such type of evolution of the universe in  brane cosmology \cite{Astashenok:2013lda}.
The solution for $\phi$ from (\ref{phi-e1}) is
\beq\label{phi-sol-2}
\phi-\phi_*=\pm \sqrt{3m}\ln t.
\eeq
It is interesting to mention that we have got the same solution for $\phi $ as in the case of power-law evolution. The reason is that the Hubble parameters $H$ defer by the constant:
$$
H_{pl}=\frac{m}{t};~~
H_{pl-exp}=\frac{m}{t}+\lambda.
$$
Therefore we have the same right hand side in (\ref{phi-e1}) for finding the field $\phi$.
We note, that the acceleration is positive: $ \frac{\ddot{a}}{a}=\frac{m(m-1)}{t^2}+\frac{2m\lambda}{t}+\lambda^2>0.$
}

Evidently, from (\ref{psi-e1}), we  obtain another result because  $H_{pl-exp}$ is involved in the equation. Thus, performing
integration (\ref{psi-e1}), we find

\beq\label{psi-sol-2}
\psi-\psi_*=\pm\sqrt{6\agb m}\left( -\frac{m}{t}+\lambda\ln t \right).
\eeq
Using the obtained solution for the chiral field $\phi$, we can make the reconstruction of the potential $V_1$ from its dependence on time $t$. Thus we obtain

\beq\label{V1-sol-2}
V_1(\phi)/3=\left(2m-\Half\right)m \exp\left(\mp 2\frac{\phi-\phi_*}{\sqrt{3m}}\right)+ 4m\lambda \exp\left(\mp \frac{\phi-\phi_*}{\sqrt{3m}}\right) +2\lambda^2.
\eeq
For the second field $\psi$ it is impossible to make a reconstruction to get the explicit dependence $V_2$ on $\psi$. Therefore we can present the expression for $V_2$ in terms of $t$

\beq\label{V2-sol-2t}
V_2=6\agb \left(\frac{m}{t}+\lambda \right)^2 \left( \frac{m(m-1/2)}{t^2}+2 \frac{m\lambda}{t}+\lambda^2\right).
\eeq

With the help of equation (\ref{psi-sol-2}) we can express $ m/t$ in terms of $\psi$ and $ \ln t$
$$
m/t=F(\psi,t)=\lambda \ln t \mp \frac{\psi}{\sqrt{6\agb m}}.
$$
Then after substituting $F(\psi,t)$ into (\ref{V2-sol-2t}), one can obtain

\beq\label{V2-psi-lnt}
V_2(\psi, t)= 6\agb \left( F(\psi,t) +\lambda\right)^2 \left(\left(1-\frac{1}{2m}\right) F(\psi,t)^2 +2\lambda F(\psi,t) +\lambda^2 \right).
\eeq

\subsection{Sin-Hyperbolic evolution}

As an example of hyperbolic evolution of the Universe let us consider the scale factor

\beq\label{sf-sht}
a(t)=A \sinh \lambda t,~~\lambda=const.>0.
\eeq
Such type of evolution was studied in \cite{Ellis:1990wsa}
The solution for $\phi$ from (\ref{phi-e1}) can be written as
\beq\label{phi-sol-3}
\phi-\phi_*=\pm \Half\sqrt{3}\ln \left( \frac{\cosh \lambda t -1}{\cosh \lambda t +1}\right).
\eeq
what is equivalent to
\beq\label{phi-sol-3a}
\phi-\phi_*=\pm \sqrt{3}\ln \left( \tanh \left(\frac{\lambda t}{2}\right) \right).
\eeq
The second field $\psi$ can be defined by integrating (\ref{psi-e1}) and it gives

\beq\label{psi-sol-3}
\psi-\psi_*=\mp \frac{\lambda\sqrt{6\agb}}{\sinh \lambda t}.
\eeq
Using the obtained solutions for the chiral fields $\phi$ and $\psi$, we can reconstruct the potentials $V_1$ and $V_2$ from their dependence on time $t$. Finally we obtain
\beq\label{V1-sol-3}
V_1(\phi)=\lambda^2 \left(6+4.5 \sinh^2 \frac{(\phi-\phi_*)}{\sqrt{3}}\right),
\eeq
\beq\label{V2-sol-3}
V_2(\psi)=6\agb \lambda^4\left(1+\frac{(\psi-\psi_*)^2}{6\agb \lambda^2}\right)
\left(1+\Half\frac{(\psi-\psi_*)^2}{6\agb \lambda^2}\right).
\eeq
Considering the very early stages of the  evolution of the universe, we let time $t$ to zero:$~ t\rightarrow 0$. Then, it is easy to see that both fields tend to infinity: $ \phi , \psi \rightarrow \mp \infty$. The same will be true for the potential: $V=V_1+V_2,~V \rightarrow \infty $.

\subsection{Cos-Hyperbolic evolution}

 We consider nonsingular evolution of the universe with the scale factor

\beq\label{sf-cht-1}
a(t)=A \cosh \lambda t,~~\lambda=const.>0.
\eeq
We note, that we have once again accelerated expansion with $\ddot{a}=A\lambda^2 \cosh \lambda t >0 $.

In the work \cite{Ellis:1990wsa}, such type of evolution was considered, but it was necessary to add the dust part to avoid imaginary values of a scalar field. In our case, for the evolution (\ref{sf-cht-1}), $  \dot{H}>0$, and we can see from (\ref{phi-e1})-(\ref{psi-e1}) that
if we want to work with real chiral fields, we have to choose the negative sign for $h_{11}$ and $h_{22}$, i.e.,  both fields should be phantom ones.  Let $h_{11}=-1$ and $h_{22}=-1$ . Taking into account these relations, one
can find the solution for $\phi$ from (\ref{phi-e1})

\beq\label{phi-sol-3}
\phi-\phi_*=\pm 2\sqrt{3}\arctan\left( e^{\lambda t}\right).
\eeq

The second field $\psi$ can be defined by integrating (\ref{psi-e1}), and it gives

\beq\label{psi-sol-3}
\psi-\psi_*=\mp \frac{\lambda\sqrt{6\agb}}{\cosh \lambda t}.
\eeq
Using the obtained solutions for the chiral fields $\phi$ and $\psi$, we can reconstruct the potentials $V_1$ and $V_2$ from their dependence on time $t$. Finally we obtain
\beq\label{V1-sol-3}
V_1(\phi)=\frac{3}{2}\lambda^2 \left(3\cos^2 \frac{(\phi-\phi_*)}{\sqrt{3}}+1\right),
\eeq
\beq\label{V2-sol-3}
V_2(\psi)=6\agb \lambda^4\left(1-\frac{(\psi-\psi_*)^2}{6\agb \lambda^2}\right)
\left(1-\Half\frac{(\psi-\psi_*)^2}{6\agb \lambda^2}\right).
\eeq
In this case the fields and the potential will tend to constants in the very early stages of the evolution of the universe when $~ t\rightarrow 0$.

It is worthwhile  to mention that the analog of such a solution for the model with kinetic interaction contradicts  the suggested decomposition or may contain imaginary fields.

\section{Exact solutions with kinetic interaction between fields}

In this section we give examples when  exact solutions exist with kinetic interaction between scalar fields. We recall that it was  impossibile to get  exact solutions in the emergent universe in EGB gravity as suggested in \cite{Chervon:2014tra} and under that decomposition.

\subsection{Power-law expansion}

For the power-law expansion (\ref{a-powlaw}), by suggesting
\beq\label{h22-pl}
h_{22}=Ct^{2m-1},~~C=const.
\eeq
we obtain the solution for the chiral fields \beq
\phi-\phi_*=\sqrt{3m}\ln t,~~ \psi-\psi_*=\sqrt{6\agb
\frac{m^3}{C}}\left(-\frac{2}{2m+1}\right)t^{-m-\frac{1}{2}},~~m
\ne -1/2.
\eeq
The reconstruction of potentials gives
\beq
V_1(\phi)=m\left( 6m-3/2\right)\exp \left(
-2\frac{\phi-\phi_*}{\sqrt{3m}}\right),
\eeq

\beq 
V_2(\psi)=3\agb m^3(2m-1)\left(
\frac{(2m+1)\sqrt{C}}{2\sqrt{6\agb
m^3}}(\psi-\psi_*)\right)^{\frac{5}{m+\frac{1}{2}}}.
\eeq 
Thus for
the power-law kinetic interaction (\ref{h22-pl}), we found the
chiral fields to  be dynamic and both parts of the potential.

\subsection{Power-law-exponential expansion}

The scale factor for power-law-exponential inflation has the form:
\begin{equation}\label{12}
a(t)=At^m e^{\lambda t}.
\end{equation}
Kinetic and potential energy can be written in this case as follows:
\begin{equation}\label{13}
K(t)=\frac{3}{2}\frac{m}{t^2}+3\agb\frac{m}{t^2}\left(\frac{m^2}{t^2}+
2\lambda\frac{m}{t}+\lambda^2\right),
\end{equation}
\begin{equation}\label{14}
V(t)=6\left(\frac{m}{t}+\lambda\right)^2-\frac{3}{2}\frac{m}{t^2}+
6\agb\left(\frac{m}{t}+\lambda\right)^2\left(\left(\frac{m}{t}+
\lambda\right)^2-\frac{1}{2}\frac{m}{t^2}\right).
\end{equation}
In the model under consideration, we can obtain the exact solution
for power-law-exponential evolution using the ansatz
(\ref{antz-1}) and the relation (\ref{f-h22}). We choose the term
responsible for kinetic interaction of the fields  to have the
form
\beq
h_{22}(\phi)=
2\frac{\lambda}{m^2}\left(m\exp\left(-\frac{\phi-\phi_*}{\pm\sqrt{3m}}\right)+
\lambda\right)^2. %~~\tilde{\phi}=\frac{\phi-\phi_*}{\pm\sqrt{3m}}
\eeq
The solution for chiral fields dynamic described by %
\begin{equation}\label{15}
\phi-\phi_*=\pm \sqrt{3m}\ln t,\,\psi-\psi_*=\pm\sqrt{3\agb
m}\frac{m}{\sqrt{\lambda}}\ln t .
\end{equation}
Below we will use the following notation
$$
\tilde{\phi}=\frac{\phi-\phi_*}{\pm\sqrt{3m}}, %~~\tilde{\psi}=\frac{\sqrt{\lambda}\psi}{\pm\sqrt{3\agbm}m}
~~\tilde{\psi}=\pm\sqrt{\frac{\lambda}{3\agb m}}\frac{\psi-\psi_*}{m}.
$$

The potential's components are 
\begin{equation}\label{16}
V_1(\phi)=6\left(m e^{-\tilde{\phi}}+\lambda\right)^2-
\frac{3}{2}e^{-\tilde{\phi}},
\end{equation}
\begin{equation}\label{17}
V_2(\psi)=6\agb\left(m e^{-\tilde{\psi}}+
\lambda\right)^2\left(\left(m e^{-\tilde{\psi}}+
\lambda\right)^2-\frac{1}{2}m e^{-\tilde{\psi}}\right)e^{-f(\psi)},
\end{equation}
\begin{multline}\label{18}
e^{f(\psi)}=\left(\frac{e^{-\tilde{\psi}}+
\frac{\lambda}{m}}{\left(3e^{-2\tilde{\psi}}+
4e^{-\tilde{\psi}}\frac{\lambda}{m}+
\left(\frac{\lambda}{m}\right)^2\right)^{1/3}}\right)^{2/m},%\\
%\times
\exp\left[-\frac{2\sqrt{2}}{3m}\arctan\left(\frac{m}{\lambda 2\sqrt{2}}\left(6e^{-\tilde{\psi}}
+4\frac{\lambda}{m}\right)\right)\right],
\end{multline}
\begin{multline}\label{19}
e^{f(\phi)}=\left(\frac{e^{-\tilde{\phi}}+
\frac{\lambda}{m}}{\left(3e^{-2\tilde{\phi}}+4e^{-\tilde{\phi}}\frac{\lambda}{m}
+2\left(\frac{\lambda}{m}\right)^2\right)^{1/3}}\right)^{2/m},%\\
%\times
\exp\left[-\frac{2\sqrt{2}}{3m}\arctan\left(\frac{m}{\lambda 2\sqrt{2}}\left(6e^{-\tilde{\phi}}
+4\frac{\lambda}{m}\right)\right)\right],
\end{multline}
\begin{equation}\label{20}
V(\phi,\psi)=V_1(\phi)+e^{f(\phi)}V_{2}(\psi).
\end{equation}

In the formulae above $f(\phi)$ and $f(\psi)$ have the same representation in terms of time $t$. Let us note that, to obtain $V(\phi,\psi)$ in the correct form (54), we must  express the three functions $V_1, e^{\phi}$ and $V_2$ as  functions of their arguments, $\phi, \phi $ and $\psi $, respectively.  Therefore, for example, to express the function $V_2$ as a function on $t$ we will use the dependence $V_2$ on $t$ (from (47)) and then, defining the cosmic time $t$ as the function on $\psi$ from the second formula in (49), we insert the result in $V_2(t)$ to obtain the desired dependence $V_2(\psi)$.

\subsection{Sin-Hyperbolical expansion}

We choose the  scale factor as
\begin{equation}\label{25}
a(t)=A\sinh(\lambda t).
\end{equation}
The dynamic chiral fields  are described by

\begin{equation}\label{28}
\phi-\phi_*=\pm2\sqrt{3} \tanh^{-1}\left(e^{\lambda t}\right),
\end{equation}

\begin{equation}\label{29}
\psi-\psi_*=\pm\sqrt{6\agb}\lambda\ln(\sinh(\lambda t)).
\end{equation}
We introduce additional functions
\begin{equation}\label{30}
\Theta(\phi)=\ln\left(\tanh\left(\exp\left(\pm\frac{\phi-\phi_*}{2\sqrt{3}}\right)\right)\right)
\end{equation}
and
\begin{equation}\label{31}
\chi(\psi)=\sinh^{-1}\left(\exp\left(\pm\frac{\psi-\psi_*}{\lambda\sqrt{6\agb}}\right)\right).
\end{equation}
Then the solution takes the form
\begin{equation}\label{32}
V_1(\phi)=\frac{9\lambda^2}{2}\sinh^{-2}\Theta (\phi)+6\lambda^2,
\end{equation}

\begin{equation}\label{33}
f(\phi)=\left(\frac{3}{\sinh^2(\Theta(\phi))}+2\right)^{1/3},
\end{equation}

\begin{equation}\label{34}
V_2(\psi)=3\bar{\alpha}\lambda^4\frac{\cosh^2(\chi(\psi))
(\cosh^2(\chi(\psi))+1)}{(2\cosh^2(\chi(\psi))+1)^{1/3}}\sinh^{-5/3}(\chi(\psi)).
\end{equation}

Thus we proved that the singular solution of sinh-type may be presented by the model with kinetic interaction of fields.

\section{Discussions}

We considered power-law, power-law-exponential and hyperbolic types for the scale factor  in EGB cosmology with a two-field CCM. First of all, we found the exact solutions for scalar fields without kinetic interaction in analogy with the emergent universe scenario \cite{Chervon:2014tra}. The decomposition used in that work seemed to indicate that solutions with kinetic interaction are impossible. Nevertheless, when we extended our approach for scale factor evolution different from that of the emergent universe scenario, solutions with kinetic interaction were found.

If we can find a reasonable way of compactification of the 5th dimension, and make sure that spectral  parameters can be calculated without perturbation analysis, then the exact solutions we found  can be confronted with observational data. We were not looking for any relation with $f(R)$ theories, but note that a connection with $R^2$ theory is important if one wishes to study inflation,  but these are topics for future investigation.

Perhaps  dark energy could be described by some of our solutions obtained.

\section{Acknowledgements}
The article is a reflection of the talk given at the Ulyanovsk International School-Seminar
"Problems of theoretical and observation cosmology - UISS 2016", September 19-30, Ulyanovsk, Russia.

SVC was partly supported by the State order of Ministry of education and science of RF number  2014/391 on the project 1670.

\end{document}